\documentclass[journal=jacsat,manuscript=article]{achemso}
\usepackage[T1]{fontenc}
\usepackage[table,xcdraw]{xcolor}
\usepackage{amsmath,amssymb}
\usepackage{hyperref}

\author{Xiaoyu Wang}
\email{xiaoyu.wang@sorbonne-universite.fr}
\affiliation{Sorbonne Universit{\'e}, CNRS, Laboratoire de Chimie Th{\'e}orique, LCT, 75005 Paris, France}

\author{Miriam Marqu\'es}
\affiliation{CSEC, School of Physics and Astronomy, University of Edinburgh, Edinburgh, EH9 3JZ, United Kingdom}

\author{Sergio G\'omez}
\affiliation{Departament d'Enginyeria Inform\`{a}tica i Matem\`{a}tiques, Universitat Rovira i Virgili, 43007 Tarragona, Spain}
\alsoaffiliation{ComSCIAM, Universitat Rovira i Virgili, 43007 Tarragona, Spain}

\author{Francesc Serratosa}
\affiliation{Departament d'Enginyeria Inform\`{a}tica i Matem\`{a}tiques, Universitat Rovira i Virgili, 43007 Tarragona, Spain}

\author{Eva Zurek}
\affiliation{Department of Chemistry, State University of New York at Buffalo, Buffalo, NY 14260-3000, USA}

\author{Julia Contreras-Garc\'ia}
\affiliation{Sorbonne Universit{\'e}, CNRS, Laboratoire de Chimie Th{\'e}orique, LCT, 75005 Paris, France}
\email{julia.contreras_garcia@sorbonne-universite.fr}

\title{Geometry-Based Neural-Network Prediction of Electron Localization Function Topology in Dense Hydrogen}


\begin{document}

\begin{abstract}
We develop a machine-learning framework to predict the electron localization function (ELF) of pure, dense hydrogen directly from atomic geometry, bypassing explicit electronic-structure calculations. 
Trained on first-principles data spanning multiple pressure regimes in dense fluid hydrogen, the model achieves high accuracy ($R^2 > 0.99$) and faithfully reproduces the global distribution of the ELF. 
A combined real- and reciprocal-space analysis reveals that the residual error is dominated by smooth, long-wavelength components with correlation lengths exceeding typical H--H bonding scales, and that the magnitude of these components increases systematically with pressure. 
Despite being trained exclusively on dense fluid hydrogen networks, the model transfers robustly to crystalline hydrogen configurations, preserving key features of ELF topology, including critical points and hydrogen-network connectivity. 
Taken together, these results suggest a viable route toward geometry-based, high-throughput evaluation of hydrogen-networking characteristics in both fluid and crystalline hydrogen.
\end{abstract}

\section{Introduction} \label{intro}
Since the seminal proposal by Wigner and Huntington that molecular hydrogen may transform into an atomic metallic solid under extreme compression \cite{Wigner1935_JCP}, hydrogen metallization has emerged as a central problem in dense-matter physics, where electron degeneracy, strong quantum effects, and structural complexity converge to challenge electronic-structure theory and high-pressure experiments alike \cite{Gregoryanz2020_MatterRadiatExtrem,Bonitz2024_PhysPlasmas}.
It has long been expected to exhibit exotic properties such as high-temperature superconductivity \cite{Ashcroft1968_PRL} and plays a central role in models of giant-planet interiors and magnetic-field generation \cite{Guillot2005_AnnuRevEarthPlanetSci}.
In the solid state, the pressure evolution of hydrogen phases has been investigated extensively through experiments \cite{Mao1994_RevModPhys,Howie2012_PRL,Dalladay2016_Nature,Eremets2019_NatPhys,Loubeyre2020_Nature,Ji2025_Nature} and theoretical simulations \cite{Pickard2007_NatPhys,Liu2012_JPCC,Mcminis2015_PRL,Monserrat2018_PRL,Monacelli2023_NatPhys}.
Despite these sustained efforts, both the metallization pressure and the microscopic mechanism driving the transition remain actively debated.

In the liquid phase, dynamic-compression and optical experiments have reported an insulator-to-conductor transition in dense hydrogen, commonly discussed in terms of a liquid–liquid transition (LLT) \cite{Nellis1996_Science,Weir1996_PRL,Loubeyre2004_HighPresRes}.
However, disparate experimental diagnostics yield conflicting signatures of the transition—ranging from abrupt to smooth changes—and infer widely different transition pressures, leaving the nature of the LLT unresolved \cite{Knudson2015_Science,Zaghoo2017_PNAS,Celliers2018_Science}.
Complementary theoretical studies employing density functional theory (DFT) and quantum Monte Carlo (QMC) simulations have sought to clarify the LLT \cite{Scandolo2003_PNAS,Delaney2006_PRL,Morales2010_PNAS,Lorenzen2010_PRB,Mazzola2018_PRL}; nonetheless, the long spatial and temporal correlations near the transition render these \textit{ab initio} approaches highly sensitive to finite-size effects, simulation length, and methodological choices, preventing a definitive characterization of the LLT \cite{Bonitz2024_PhysPlasmas}.

To overcome the intrinsic size and time limitations of \textit{ab initio} molecular dynamics, machine-learning interatomic potentials trained on first-principles data have been developed to enable large-scale simulations of dense liquid hydrogen \cite{Cheng2020_Nature,Tirelli2022_PRB}.
These models are trained on reference datasets generated from DFT and/or QMC calculations and are constructed to faithfully reproduce \textit{ab initio} energies, forces, and stresses.
As a result, they permit systematic exploration of thermodynamic observables and structural order parameters across broad regions of the phase diagram \cite{Cheng2020_Nature,Dong2025_PRB}, enable accurate determination of pressures and equations of state \cite{Tirelli2022_PRB}, and provide access to free-energy landscapes, phase stability, and transport properties \cite{Bischoff2024_arxiv,Istas2025_PRE,Dong2025_PRB,Tenti2025_PRB}.
Collectively, these studies demonstrate that extending simulations to sufficiently large system sizes and long time scales is essential for disentangling genuine thermodynamic behavior from sampling and convergence effects in the liquid–liquid transition.

A notable limitation of current machine-learning frameworks is the lack of direct access to electronic bonding information.
In dense liquid hydrogen, molecular character is increasingly understood as a dynamic and statistical motif rather than a collection of well-defined chemical species: ab initio studies show that H–H correlations become short-lived across the LLT and that molecular fractions evolve smoothly and remain estimator-dependent \cite{Morales2010_PNAS,Bonitz2024_PhysPlasmas}.
Consequently, short H–H distances alone do not uniquely imply stable molecular bonding in this regime.
Electronic descriptors offer a complementary route to characterizing bonding beyond purely geometric criteria, most notably the electron localization function (ELF) \cite{Becke1990_JCP}, which directly quantifies electronic localization and pairing \cite{Hua2019_PRB}.
In parallel, lifetime- and cutoff-sensitive bonding diagnostics have been applied to fluid and superionic hydrogen-rich systems, where the emergence of extended hydrogen networks depends sensitively on bonding criteria and observability thresholds \cite{deVilla2025_JCP}.
While bonding analysis based on electronic information is therefore essential for organizing dissociation and metallization regimes, its reliance on orbital-level quantities in \textit{ab initio} approaches makes such descriptors computationally demanding, motivating geometry-based strategies for efficient prediction of electronic bonding measures.

The goal of the present work is to address this bottleneck by combining ab initio molecular dynamics (AIMD) data for compressed liquid hydrogen with machine-learning predictions of the electron localization function (ELF) topology based solely on local geometric descriptors, without explicit reliance on electronic wavefunctions or orbitals.
We develop a neural-network model capable of predicting the full three-dimensional ELF field with high fidelity across a wide pressure range, enabling a robust and scalable representation of the electronic localization landscape in the warm dense regime.
A systematic analysis of the residual between the predicted and reference ELF fields reveals that the remaining errors are dominated by smooth, long-wavelength components, which can be efficiently captured using a band-limited Fourier representation. This separation provides a transparent characterization of pressure-dependent nonlocal contributions and offers physical insight into the emergence of long-range correlations.
Beyond field-level prediction, the framework enables the identification of ELF critical points and the evaluation of hydrogen-networking descriptors \cite{Belli2021_NatCommun}, providing direct access to topology-based measures of bonding organization in dense hydrogen and opening avenues for future applications to more complex hydrogen-containing systems.

\section{Methodology}
\subsection{Representations}
The electron localization function (ELF) \cite{Becke1990_JCP} is typically evaluated and visualized on a real-space grid, reflecting its interpretation as a spatially resolved electronic descriptor. 
To predict the ELF on this grid, we represent the local atomic environment around each grid point ($v$) by a smooth neighbor density constructed from atomic positions and expanded in a rotation-invariant basis.
For each grid point located at fractional coordinate $\mathbf{r}_v$, we define a hydrogen-only neighbor density
\begin{equation}
    \rho(\mathbf{r};\mathbf{r}_v) = \sum_{i} w(r_{vi})\delta(\mathbf{r}-\mathbf{r}_{vi}),
\end{equation}
where $\mathbf{r}_i$ denotes the position of hydrogen atom $i$, $\mathbf{r}_{vi}=\mathbf{r}_i-\mathbf{r}_v$, and periodic boundary conditions are enforced.
In practice, the Dirac delta function is replaced by a set of smooth basis functions, yielding a continuous and differentiable representation suitable for numerical evaluation.
The weighting function
\begin{equation}
    w(r)=\frac{1}{2}[\cos(\pi r/r_\textrm{cut})+1],~r\leqslant r_\textrm{cut},
\end{equation}
smoothly truncates the density at a finite cutoff radius $r_\mathrm{cut}$, ensuring continuity of both the density and its first derivative at the cutoff.

The neighbor density is expanded in a product basis of Gaussian radial functions, $R_n(r)=\exp{\left[-(r-\mu_n)^2/(2\sigma^2)\right]}$ with centers $\mu_n$ uniformly spanning $[0,r_\textrm{cut}]$, and real spherical harmonics, $Y_{lm}$ up to angular momentum $l_\textrm{max}$
\begin{equation}
    \rho(\mathbf{r};\mathbf{r}_v) \rightarrow c_{nlm}(\mathbf{r}_v)=\sum_iR_n(r_{vi})w(r_{vi})Y_{lm}(\mathbf{\hat{r}}_{vi}).
\end{equation}

To obtain descriptors invariant under global rotations, we compute the power spectrum of the expansion coefficients
\begin{equation}
    P_{nn^\prime}^{(l)}(\mathbf{r}_v)=\sum_{m=-l}^lc_{nlm}(\mathbf{r}_v)c_{n^\prime lm}(\mathbf{r}_v)
\end{equation}
and retain only the upper-triangular components $n\leqslant n^\prime$.
The final feature vector at each grid point is the concatenation of $P_{nn^\prime}^{(l)}$ over all $l=0,\ldots,l_\textrm{max}$, resulting in a fixed-length, rotation-invariant representation:
\begin{equation}
    \mathbf{x}(\mathbf{r}_v) = \bigoplus_{l=0}^{l_{\max}} \left\{P^{(l)}_{nn'}(\mathbf{r}_v)\right\}_{n \leqslant n'}.
\end{equation}
This construction is closely related to the Smooth Overlap of Atomic Positions (SOAP) \cite{Bartok2013_PRB} and Behler–Parrinello symmetry-function formalisms \cite{Behler2007_PRL}, both widely used in machine-learning studies of dense hydrogen \cite{Cheng2020_Nature,Tirelli2022_PRB,Dong2025_PRB,Tenti2025_PRB,Istas2025_PRE}, but differs in being evaluated on a continuous real-space grid rather than at atomic centers.

All feature vectors are standardized using statistics computed on the training set.
The target ELF values are taken directly from density-functional theory calculations and are not normalized beyond their natural $[0,1]$ range. 
A grid point-wise multilayer perceptron (MLP) regressor with sigmoid output maps the local density features to the predicted ELF value at the same grid point.

\subsection{Training Data Construction}
The training datasets used in this work were extracted from the 3000\textsuperscript{th} step of each 500-atom AIMD trajectory, at which both the atomic structure and electronic degrees of freedom were fully equilibrated.
Three representative volumes were selected, corresponding to pressures of 76.0, 115.1, and 138.5~GPa at 1500~K.
These pressure conditions sample an ensemble of hydrogen-network topologies in which the dominant local bonding motifs evolve from molecular to atomic character. 
The resulting distribution of configurations, shown in Fig.~1A, spans a regime where progressive electronic delocalization and fluid metallization have been extensively discussed\cite{Bonitz2024_PhysPlasmas}. 
Even at lower pressures, intermolecular interactions already perturb the idealized molecular limit, reducing and broadening the bond-centered ELF maximum and introducing intermediate ELF values\cite{Riffet2017_PCCP}. 
Importantly, the AIMD configurations naturally span the ELF regimes that govern the networking analysis: the networking value is determined by connectivity transitions at intermediate ELF isovalues rather than by regions where ELF $\approx 1$\cite{Belli2021_NatCommun,Novoa2025_ChemSci}.

The machine-learning training set was constructed by combining the ELFs calculated for three independent MD snapshots, obtained at the $P-T$ conditions noted, and calculated on a $192^3$ real-space grid. In each case, 50,000 grid points, and therefore local atomic environments, were randomly chosen using a stratified scheme that is approximately uniform in ELF value, yielding a total training set of 150,000 training points.
Structural descriptors were generated within a radial cutoff of 3.0~\r{A}, using 10 radial basis functions and spherical harmonics up to $l_{\max}=2$.
All features were stored in half precision (float16) to reduce memory footprint. 
Sampling was performed in batches of 1024 with a fixed random seed to ensure reproducibility. 

It is worth noting that all reference calculations in this study are performed using the PBE exchange--correlation functional. Consequently, the predicted ELF inherits the known limitations of this approximation, and the achievable model accuracy is fundamentally constrained by the fidelity of the underlying electronic structure description.
While different functionals may shift the pressure at which the molecular-to-network transition occurs, the associated evolution of ELF topology and connectivity is expected to remain qualitatively robust.

\subsection{Model Architecture and Training Protocol}

We employ a multilayer perceptron (MLP) regressor \cite{Hornik1989_NeuralNetw} consisting of two hidden layers of width 128 with smooth rectified linear unit (SiLU) activations, implemented in PyTorch \cite{Paszke2019_NIPS}. 
The model maps the descriptor vector $\mathbf{X}\in\mathbb{R}^{F}$ to a scalar target via a fully connected architecture.
The network is trained for 80 epochs using the AdamW optimizer \cite{Loshchilov2017_arxiv} with learning rate $3\times10^{-4}$ and weight decay $10^{-6}$. 
Optimization minimizes the Huber loss \cite{Huber1992} with parameter $\beta=0.03$, which provides a quadratic penalty for small residuals and a linear penalty for large residuals, improving robustness to outliers while retaining sensitivity near the optimum.

Training is performed with a batch size of 4096. All input features are standardized to zero mean and unit variance using statistics computed on the training set only, thereby preventing information leakage into validation data. Model parameters are initialized with a fixed random seed to ensure reproducibility.
Hyperparameters—including network width, depth, cutoff radius $r_{\mathrm{cut}}$, number of radial basis functions $n_{\mathrm{radial}}$, maximum angular momentum $l_{\max}$, and the Huber loss parameter—are systematically optimized via controlled one-factor-at-a-time scans, as detailed in Section~S2.

\begin{figure*}[!ht]
    \centering
    \includegraphics[width=0.85\linewidth]{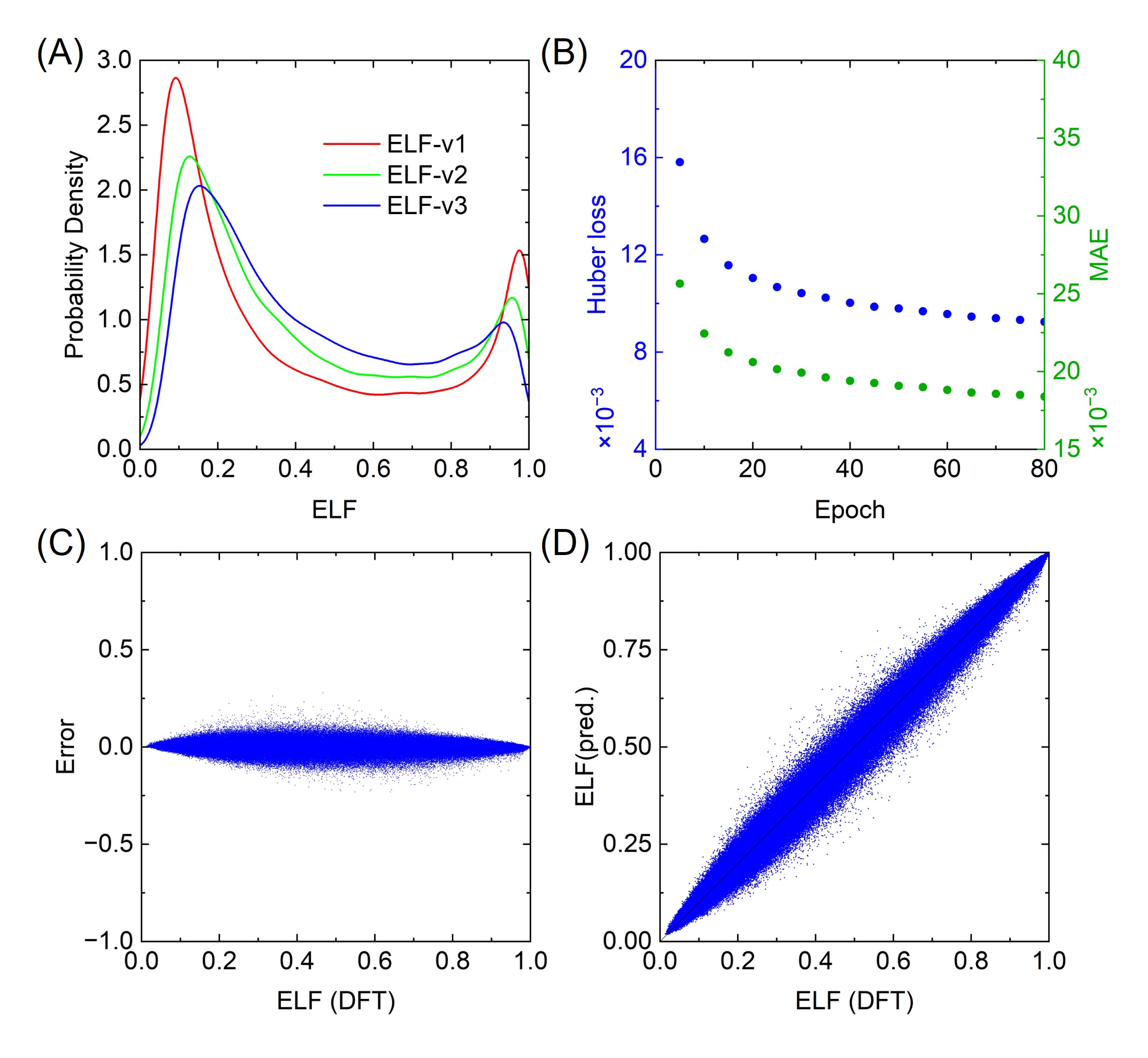}
\caption{
Data distribution, training convergence, and predictive performance of the ELF model.
(A) Probability density distribution of ELF values across the dataset, illustrating the broad sampling of the target space.
(B) Training Huber loss and validation mean absolute error (MAE) as a function of training epoch.
(C) Prediction error as a function of the reference DFT ELF value.
(D) Predicted versus reference DFT ELF values, with the solid line indicating the identity.
}
\label{fig:1}
\end{figure*}

\section{Results and discussion}

\subsection{Performance}
The model was trained for 80 epochs, after which both the training Huber loss and the validation mean absolute error (MAE) reach clear plateaus, indicating stable convergence without evidence of overfitting (Figure~\ref{fig:1}B).
The concurrent saturation of these metrics suggests that the representational capacity of the local-descriptor model is fully exploited within this training window.

The validation set was generated using the same sampling protocol as the training data, however in this case 150,000 local atomic environments were extracted from the dataset at each of the three pressure conditions.
Figures~\ref{fig:1}C and \ref{fig:1}D summarize the predictive performance over the training set. 
The error distribution as a function of the reference ELF (Figure~\ref{fig:1}C) is narrowly centered around zero across the full ELF range, with no discernible systematic bias at either low or high localization values. 
Correspondingly, the predicted ELF values closely follow the identity line (Figure~\ref{fig:1}D), yielding excellent agreement with the DFT reference statistics 
(MAE = $0.0190$ , root mean squared error (RMSE) = $0.0271$, and coefficient of determination ($R^2$) = $0.992$).
The predicted and reference distributions are nearly indistinguishable, with identical means (0.426) and very similar variances, confirming that the model accurately reproduces both the central tendency and overall spread of ELF values.
The stability of the model was further evaluated by repeating the entire workflow (including dataset construction, training, and validation) 50 times with different random seeds. 
Although trained on the 3000th snapshot of each AIMD run, the model achieves comparable accuracy on the 1000th and 2000th snapshots used as an independent validation set (Table~S2).
The results demonstrate high reproducibility, with negligible variation in all reported metrics. Further details are provided in Section S3 of the Supporting Information.

Despite the high overall fidelity, the residual field is not purely stochastic.
As shown in Fig.~\ref{fig:1}C, the prediction error varies across the ELF range, with larger absolute deviations observed at intermediate ELF values and a correspondingly broader spread around the identity line in Fig.~\ref{fig:1}D.
This region corresponds to electronic environments with ELF values close to that of a homogeneous electron gas (HEG, ELF $\approx 0.5$), where electron localization is weak and bonding characteristics are less clearly defined.
In this regime, relatively small geometric or electronic variations can lead to noticeable changes in ELF, making prediction more challenging.
In contrast, highly localized (ELF $\rightarrow 1$) and strongly depleted (ELF $\rightarrow 0$) regions exhibit more distinct signatures and are predicted with higher accuracy.
The enhanced deviations at intermediate ELF values therefore motivate a more detailed analysis of the structured residuals, which is presented in the following section.

\subsection{Origin of the Residual}
\begin{figure*}[!htbp]
    \centering
    \includegraphics[width=1.0\linewidth]{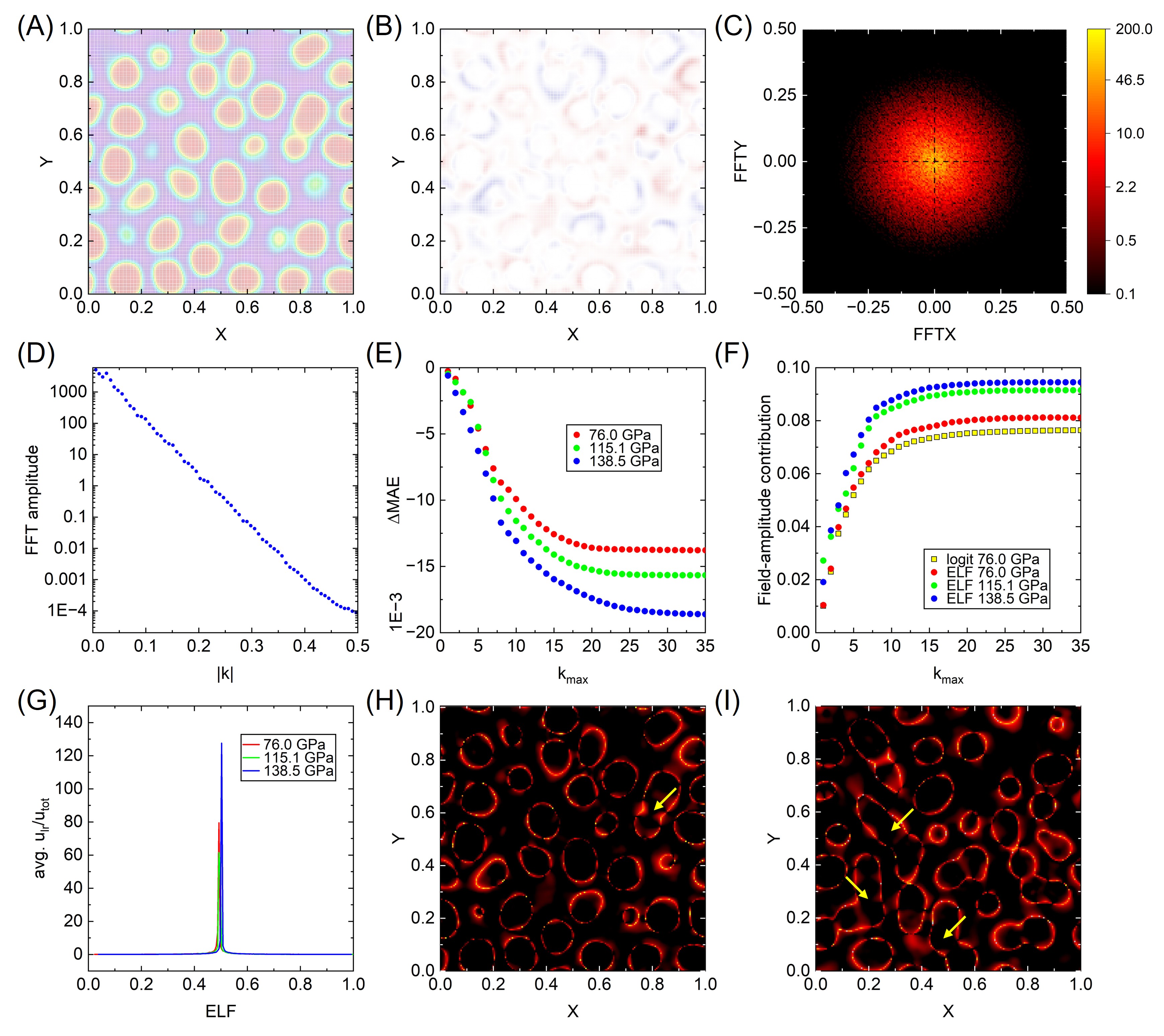}
    \caption{ELF and residual-field analysis for a representative two-dimensional slice of the 76.0~GPa structure.
    (A) DFT-computed ELF in real space (color scale from 0 (blue) to 1 (red); fractional coordinates relative to 11~\AA).
    (B) Real-space residual between predicted and DFT ELF (color scale from $-0.2$ (blue) through 0 (white) to $+0.2$ (red)).
    (C) Two-dimensional Fourier transform of the residual field.
    (D) Radially averaged Fourier amplitude as a function of wave vector $k$.
    (E) Change in mean absolute error ($\Delta$MAE) versus Fourier cutoff $k_\mathrm{max}$ for 76.0, 115.1, and 138.5~GPa.
    (F) Fractional magnitude of the Fourier correction versus $k_\mathrm{max}$, evaluated in logit space at 76.0~GPa (yellow squares) and in ELF space at 76.0, 115.1, and 138.5~GPa (circles).
    (G) Distribution of the Fourier-correction contribution ($u_\mathrm{lr}/u_\mathrm{tot}$) as a function of ELF.
    (H) Real-space distribution of the Fourier-correction contribution at 76.0 and (I) 138.5~GPa (logarithmic color scale from 0.1 (black) to 25(yellow)).}
    \label{fig:2}
\end{figure*}

To clarify the physical origin of the residual field not captured by the local descriptor model, we examine its real-space structure using a representative two-dimensional slice from the 76.0~GPa configuration (Fig.~\ref{fig:2}A).
The corresponding residual field (Fig.~\ref{fig:2}B) exhibits a weak, smoothly varying modulation extending over a substantial fraction of the unit cell.
The residual magnitude is uniformly small, with most values confined within 0.05 and only rare, spatially diffuse regions approaching larger deviations.
These features span several tenths of the unit cell, corresponding to correlation lengths of a few angstroms, well above typical bond lengths, indicating a low-amplitude, long-wavelength contribution rather than missing short-range or chemically specific effects.

This picture is reinforced by reciprocal-space analysis.
The two-dimensional Fourier spectrum of the residual field (Fig.~\ref{fig:2}C) is strongly $\Gamma$-centered and nearly isotropic, with spectral weight concentrated at small wavevectors and smoothly decaying toward the Brillouin-zone boundary.
No anisotropy, lattice-locked features, or enhancement at large $\lvert\mathbf{k}\rvert$ are observed.
Consistently, the radially averaged Fourier amplitude (Fig.~\ref{fig:2}D) decreases monotonically with increasing $\lvert\mathbf{k}\rvert$, without secondary peaks or characteristic length scales.
These observations demonstrate that the residual is dominated by coherent long-wavelength contributions not captured by strictly local descriptors, establishing a clear separation between local and collective components of the ELF.

Motivated by this clear separation of length scales, the residual can be systematically reduced by introducing a HEG-like background as a long-range correction. 
The correction is naturally formulated in the logit representation of the ELF, which provides an unbounded scalar field suitable for additive decomposition.
The real-space logit associated with the locally predicted ELF, $y_\mathrm{loc}$, is defined as
\begin{equation}
u_\mathrm{loc}(\mathbf{r})
= \mathrm{logit}\!\left(y_\mathrm{loc}(\mathbf{r})\right)
= \ln\!\left[\frac{y_\mathrm{loc}(\mathbf{r})}{1-y_\mathrm{loc}(\mathbf{r})}\right],
\end{equation}
where $\mathbf{r}$ denotes the fractional coordinate within the unit cell.
Within this representation, long-range contributions, $u_\mathrm{lr}(\mathbf{r})$, are expressed as a smooth, band-limited Fourier expansion as,
\begin{equation}
u_\mathrm{lr}(\mathbf{r}) = \sum_{|\mathbf{k}|<k_\mathrm{cut}}
\left[
  a_{\mathbf{k}} \cos\!\left(2\pi \mathbf{k}\!\cdot\!\mathbf{r}\right)
  + b_{\mathbf{k}} \sin\!\left(2\pi \mathbf{k}\!\cdot\!\mathbf{r}\right)
\right],
\end{equation}
where $k_\mathrm{cut}$ sets the maximum spatial frequency retained in the long-range field, and the coefficients $a_{\mathbf{k}}$ and $b_{\mathbf{k}}$ are determined by a least-squares fit to the residual.
By construction, this expansion captures only smooth, collective variations while excluding short-range, atom-centered features already described by $u_\mathrm{loc}$. 

The total logit field is obtained through the additive decomposition
\begin{equation}
u_\mathrm{tot}(\mathbf{r})
= u_\mathrm{loc}(\mathbf{r}) + u_\mathrm{lr}(\mathbf{r}),
\end{equation}
and the corrected ELF is recovered by mapping back to the physical interval $[0,1]$ using the inverse logit transformation:
\begin{equation}
y_\mathrm{pred}(\mathbf{r})
= \frac{1}{1 + \exp\!\left[-u_\mathrm{tot}(\mathbf{r})\right]}.
\end{equation}

We emphasize that this long-range correction is not introduced as a practical solution for improving transferable ELF predictions. 
Rather, it serves as an interpretive tool that exposes the separation between short-range contributions captured by strictly local geometric descriptors and residual long-wavelength components that reflect collective, nonlocal electronic effects. 
By isolating these smooth contributions in a controlled manner, the analysis provides physical insight into the nature and pressure evolution of nonlocal correlations, rather than constituting an additional predictive model.

Taking the two-dimensional slice discussed above as a representative example, the analysis is performed on a $192\times192$ real-space grid extracted from the 76.0~GPa structure.
For this slice, the purely local-descriptor baseline yields an MAE of 0.0184 (RMSE $\approx 0.0283$).
Figure~\ref{fig:2}E shows the evolution of the MAE reduction, $\Delta$MAE, as a function of the maximum Fourier cutoff $k_\mathrm{max}$ used to construct the long-range correction in logit space.
As $k_\mathrm{max}$ increases, $\Delta$MAE decreases monotonically and converges rapidly, reaching a clear plateau at $k_\mathrm{max}\approx 20$.
Beyond this cutoff, no further improvement is observed, and the MAE reduction saturates at $\Delta\mathrm{MAE}\approx -0.0137$ ($\Delta\mathrm{RMSE}\approx -0.022$), corresponding to a remaining error of MAE $\approx 0.0047$ and RMSE $\approx 0.0063$.
The progressive improvement of the residual maps with increasing $k_\mathrm{max}$ is shown in Figure~S3.
This convergence behavior demonstrates that the residual error of the local-descriptor model is almost entirely accounted for by a band-limited, long-wavelength field.

The quantitative contribution of the long-range correction is summarized in Fig.~\ref{fig:2}F.
Despite its pronounced impact on prediction accuracy, the long-range component contributes only a small fraction of the total field amplitude: approximately 7.6\% of the total logit field $u_\mathrm{tot}$ and 8.1\% of the corrected ELF field $y_\mathrm{pred}$ at 76.0~GPa.
This apparent disparity highlights an important physical point: although small in magnitude, the coherent low-$k$ structure of the long-range component enables it to correct systematic errors that are inaccessible to strictly local descriptors.
The residual therefore represents a genuinely nonlocal contribution—minor in amplitude, yet essential for quantitative accuracy.

A clear pressure dependence is observed in the magnitude and spatial extent of the long-wavelength residual.
The same analysis, performed on analogous two-dimensional slices at 115.1 and 
138.5~GPa,
yields baseline MAEs of $0.0192$ and $0.0197$, respectively, indicating progressively stronger long-range residuals at higher pressure.
Accordingly, convergence of the Fourier correction requires progressively higher cutoffs, with $k_\mathrm{max}\approx 24$ at 115.1~GPa and $k_\mathrm{max}\approx 27$ at 
138.5~GPa,
consistent with a more plane-wave-like character of the residual field (Figure~\ref{fig:2}E).
The fractional contribution of the correction in ELF space also increases to approximately 9.1\% and 9.4\%, respectively (Figure~\ref{fig:2}F).

We next examine the ELF- and real-space structure of this correction. 
As shown in Figure~\ref{fig:2}G, the Fourier correction is strongly concentrated around $\mathrm{ELF}\!\approx\!0.5$, indicating that the long-wavelength contribution primarily resides in electronically intermediate regions rather than at fully localized maxima or minima. 
This pressure-driven evolution is also evident in real space. Figure~\ref{fig:2}H shows the spatial distribution of the Fourier-correction contribution (corresponding to Figure~\ref{fig:2}A at 76.0~GPa). 
The red contours typically enclose a single ELF localization center, consistent with a regime dominated by localized bonds or lone-pair–like features at this pressure (and 1500~K), with polymerized hydrogen motifs occurring only rarely. 
Upon increasing the pressure to 138.5~GPa, an extended pattern of polymerization emerges across large regions of the cell: the contours connect neighboring ELF maxima, forming continuous networks that span multiple localization centers (Figure~\ref{fig:2}I). 
These contours represent the spatial weighting of the retained Fourier components, thereby highlighting electronic features that require intrinsically delocalized descriptors. 
Consistent with this interpretation, exploratory convolutional models \cite{Lecun2002_ProcIEEE} indicate that ELF prediction is dominated by short-range contributions (Section S1), and that the remaining long-wavelength residual arises from intrinsic locality limitations rather than model inadequacy.
Taken together, these observations reveal a clear mechanism for pressure-induced metallization in hydrogen networks, whereby progressive electronic delocalization manifests as the growth, compression, and interconnection of long-range correlation structures in real space.

\subsection{ELF Networking Validation}

The training data employed in this work are obtained from AIMD simulations of dense hydrogen, a setting that naturally enables detailed analysis of bonding evolution and phase transitions under high-pressure and high-temperature conditions, and within which both the model accuracy and the physical origin of the residual have been established. 
A further, more challenging step is to examine the predictive power of the model with respect to ELF topology, including the extraction of critical points in ground-state structures to quantify hydrogen-framework networking, defined as the highest ELF isovalue at which a continuous, crystal-spanning network of electronic localization—mediated by ELF saddle points—is formed \cite{Belli2021_NatCommun}.
This capability would enable rapid estimates of the superconducting critical temperature ($T_\mathrm{c}$), as implemented in the \textsc{TcEstime} framework \cite{Novoa2025_ChemSci}. 
Such a pipeline is particularly attractive as a filtering stage in crystal-structure-prediction (CSP) workflows or random-structure searches targeting superconducting hydrides \cite{Fang2025_PRB,Belli2025_AnnPhys}. 
For these use cases, performance must be evaluated on randomly generated structures to assess transferability outside the training domain.

\begin{figure*}[!htbp]
    \centering
    \includegraphics[width=1.0\linewidth]{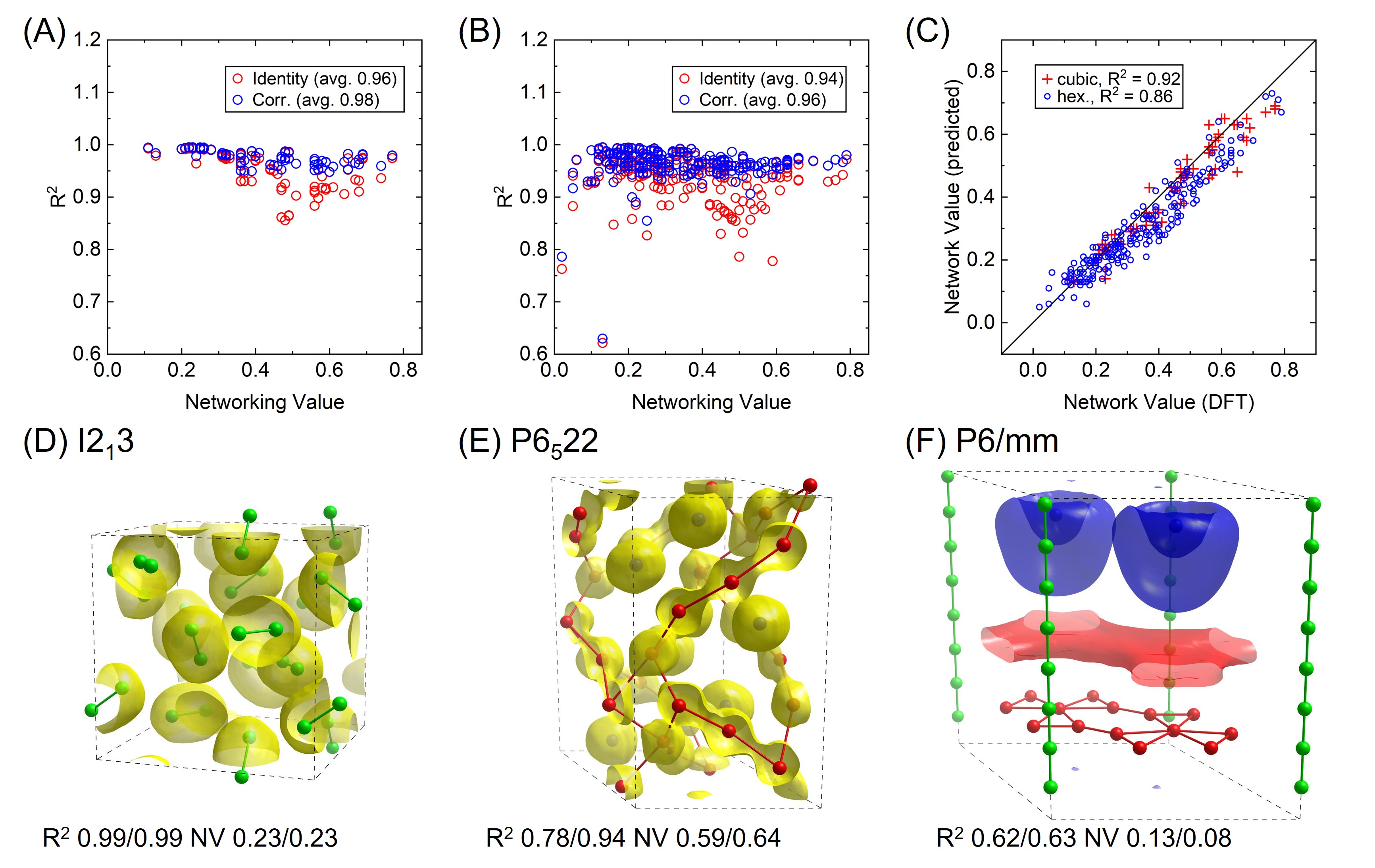}
    \caption{Performance of the local-descriptor ML model for predicting ELF-derived hydrogen networking values.
    (A,B) Coefficient of determination ($R^2$) of the pointwise ELF prediction error for individual structures, shown as a function of the networking value for (A) cubic and (B) hexagonal structures optimized at 100~GPa.
    (C) ML-predicted versus DFT networking values for cubic (red) and hexagonal (blue) systems; the solid line denotes perfect agreement.
    (D) Cubic $I2_13$ structure with isolated H$_2$ units (ELF isovalue 0.85).
    (E) Hexagonal $P6_522$ structure featuring a three-dimensional hydrogen framework and isolated H$^-$ species (ELF isovalue 0.65).
    (F) Hexagonal $P6/mm$ structure with one-dimensional hydrogen chains and extended interstitial regions, shown as $\Delta$ELF at isovalue 0.25.
    Green indicates H$_2$ molecules or polymeric chains, red the three-dimensional hydrogen framework, and blue H$^-$ species.
    The annotations in panels (D–F) are reported as ``$R^2$ (identity/correlated), $\phi$ (ML/DFT)'', where the identity $R^2$ measures agreement with the identity line, and the correlated $R^2$ measures linear correlation independent of deviations from the identity. $\phi$ denotes the networking value predicted by ML and computed from DFT, respectively.}
    \label{fig:3}
\end{figure*}

To this end, we generated cubic and hexagonal hydrogen lattices using the \textsc{RandSpg} code.\cite{Avery2017_ComputPhysCommun}
A total of 119 cubic and 230 hexagonal structures were constructed and fully optimized at 100~GPa.
Only local geometric descriptors were employed in the present model, as no transferable global long-range correction field was trained across distinct geometries.
The fitting performance for both crystal families is summarized in Fig.~\ref{fig:3}A and B, where the $R^2$ of the pointwise ELF prediction is analyzed as a function of the networking value.
Two complementary metrics are reported: the identity $R^2$ measures agreement with the identity line, and the correlated $R^2$ which measures linear correlation between predicted and DFT ELF values independent of deviations from the identity line.
A low identity $R^2$ combined with a high correlated $R^2$ therefore indicates good linearity of the prediction but a systematic deviation from the identity.
Both cubic and hexagonal structures exhibit good overall agreement, with average identity $R^2$ values of 0.96 and 0.94, respectively.
A systematic deviation between identity and correlated $R^2$ is observed near intermediate networking values ($\sim$ 0.5), consistent with the increased difficulty of accurately capturing weakly localized electronic environments, as discussed above.

The direct prediction of networking values remains robust despite the increased difficulty at intermediate ELF values, as shown in Fig.~\ref{fig:3}C, where the overall $R^2$ between predicted and DFT networking values is 0.92 for cubic structures and 0.86 for hexagonal structures.
A stability analysis was performed for ELF prediction and networking value (NW) evaluation on randomly generated structures. For each structure, 50 independent model realizations were used. The results confirm that the run-to-run variability remains small, while the variation across different structures reflects intrinsic differences in structural complexity and prediction difficulty. Detailed statistics are provided in Section S3 and in the accompanying supplementary table (xlsx file).
For molecular hydrogen structures, the networking value is small because electrons are strongly localized within intramolecular H–H bond pairs, resulting in weak electronic connectivity between neighboring molecules.
A large subset of the cubic structures considered here falls into this regime and exhibits minimal systematic prediction errors (Fig.~\ref{fig:3}A).
A representative example is the cubic $I2_13$ structure shown in Fig.~\ref{fig:3}D, which consists exclusively of isolated H$_2$ molecules.
In this case, the networking value is 0.23 in both the DFT reference and ML prediction, and both identity and correlated $R^2$ values reach 0.99.
These results indicate that the ELF in molecular phases is accurately captured by purely local geometric representations.

A more challenging regime is illustrated by the $P6_522$ structure (Fig.~\ref{fig:3}E).
In this case, a clear separation emerges between the identity and correlated metrics: the predicted ELF exhibits strong linear correlation with the DFT reference (correlated $R^2$ = 0.94) but shows a noticeable deviation from the identity relation (identity $R^2$ = 0.78), indicating a systematic offset in the absolute ELF values.
This structure features a three-dimensional, delocalized ELF network coexisting with isolated hydrogen units, giving rise to extended electronic connectivity beyond strictly local environments.
Such delocalized ELF topologies are inherently more difficult to reproduce with purely local geometric descriptors, leading to reduced agreement with the identity line despite preserved linearity.
Nevertheless, the predicted networking value (0.64) remains in close agreement with the DFT value (0.59), demonstrating that the essential topological connectivity of the hydrogen framework is correctly captured even when absolute ELF values deviate systematically.

A quick estimation of the superconducting critical temperature can be obtained within the TcEstime framework. 
For the present pure hydrogen systems, the hydrogen fraction and the hydrogen-resolved density of states at the Fermi level are trivially unity, such that the descriptor reduces directly to the networking value ($\phi$). 
Using the empirical relation proposed by Belli \textit{et al.}\cite{Belli2021_NatCommun}, we obtain estimated critical temperatures spanning from $\sim 0$--400 K across representative structures (Table~\ref{tab:tc_estimate}). 
The close agreement between ML-predicted and DFT-derived values reflects the high fidelity of the model at the level of the networking descriptor, demonstrating that the physically relevant trends for superconductivity screening are preserved within this framework.

\begin{table}[!ht]
\centering
\caption{
Estimated superconducting critical temperatures $T_c$ obtained from the networking value ($\phi$) using the TcEstime relation $T_c = 750\,\phi - 85$ K. Negative values are interpreted as non-superconducting ($T_c \approx 0$ K).}
\label{tab:tc_estimate}
\begin{tabular}{lcccc}
\hline
Structure & $\phi$ (DFT) & $\phi$ (ML) & $T_c$ (DFT) [K] & $T_c$ (ML) [K] \\
\hline
$I2_13$     & 0.23 & 0.23 & 87  & 87  \\
$P6_5 22$   & 0.59 & 0.64 & 358 & 395 \\
$P6/mm$     & 0.13 & 0.08 & 13  & -25   \\
\hline
\end{tabular}
\end{table}

The most severe failure mode is observed exclusively in a subset of hexagonal structures.
Lower symmetry enables more complex hydrogen topologies, as exemplified in Figure~\ref{fig:3}F.
In this structure, one-dimensional hydrogen chains extend along the $z$ direction, while the interchain regions are occupied by alternating layers of isolated atomic hydrogen two-dimensional hydrogen networks.
Analysis of the ELF residual reveals the origin of the error: the large interstitial regions between the hydrogen atoms and the 2D network are essentially invisible to the local descriptors, resulting in a negative ELF deviation around the H$^-$ sites and a compensating positive deviation in the interstitial region.
Consequently, both identity and correlated $R^2$ values deteriorate to 0.62–0.63.
This case highlights a genuine limitation of the present model, arising from the absence of large, electronically active interstitial regions in the dense-hydrogen training data.
In practice, however, such structures are likely to be eliminated by complementary CSP filters (e.g., energetic stability) and are therefore unlikely to survive in energy-driven structure searches \cite{Falls2020_JPCC}.

Overall, these tests demonstrate that the present ML model provides reliable predictions of hydrogen networking values across a broad range of randomly generated structures.
Although global background contributions can limit pointwise ELF accuracy in specific cases, the preserved linearity ensures that derived networking values remain quantitatively meaningful.
Beyond static structure screening, this framework could be naturally coupled with machine-learning interatomic potentials to enable large-scale molecular-dynamics simulations of dense hydrogen, in which bonding topology and hydrogen-network connectivity can be analyzed without explicit \textit{ab initio} calculations.
Beyond elemental hydrogen, it is reasonable to anticipate that this approach may extend to metal-doped hydrogen-rich compounds, where the dominant bonding topology and connectivity are primarily governed by the hydrogen sublattice, while the metal species play a secondary structural or charge-balancing role.
Together, these capabilities suggest a viable route toward alleviating a major computational bottleneck in the exploration of hydrogen-rich and multicomponent chemical spaces.\cite{Zhao2024_NatlSciRev}

\section{Conclusions}
We have demonstrated that a machine-learning model based solely on local geometric descriptors can predict the electron localization function (ELF) of dense hydrogen-rich systems with high accuracy across multiple pressure regimes, achieving $R^2 > 0.99$ while faithfully reproducing the global ELF distribution.
A systematic analysis of the residual reveals that the remaining error is not stochastic, but arises from smooth, long-wavelength components with correlation lengths exceeding typical bond distances, whose magnitude and spatial extent increase with pressure.
These nonlocal contributions are efficiently captured by a band-limited Fourier correction formulated in the logit representation, despite accounting for only a small fraction of the total field amplitude.
Importantly, ELF-derived topological descriptors—specifically hydrogen-networking values—remain robust in the presence of structured residuals, enabling reliable characterization of hydrogen bonding topology across a wide range of configurations.
While the present study focuses on elemental hydrogen, these observations suggest that similar approaches may be applicable to hydrogen-rich compounds in which the dominant bonding connectivity is governed by the hydrogen sublattice, motivating future investigations in more complex chemical environments.

A key advantage of the present framework is that it completely bypasses the explicit calculation of Kohn–Sham orbitals or kinetic-energy densities, which are traditionally required for ELF evaluation.
Instead, ELF is inferred directly from atomic geometry, yielding orders-of-magnitude reductions in computational cost relative to conventional first-principles workflows.
Earlier attempts to approximate ELF using density-only formulations have been shown to suffer from limited accuracy and sensitivity to the choice of reference frame, restricting their practical applicability \cite{Tsirelson2002_ChemPhysLett}.
By contrast, the present approach achieves high fidelity without recourse to wavefunction-level information, making it well suited for high-throughput and exploratory studies where direct electronic-structure calculations would be computationally prohibitive.

\section{Computational method} \label{theo}
\subsection{\textit{Ab initio} calculations}
AIMD data were obtained from simulations reported previously \cite{Hua2019_PRB}, performed in cubic supercells containing 500 hydrogen atoms.
Electronic-structure calculations were carried out within density functional theory using the Perdew–Burke–Ernzerhof (PBE) exchange–correlation functional \cite{Perdew1996_PRL}, together with the projector–augmented-wave (PAW) method \cite{Blochl1994_PRB}, as implemented in the Vienna \textit{ab initio} Simulation Package (\textsc{VASP}) \cite{Kresse1993_PRB}.
A plane-wave kinetic-energy cutoff of 700~eV was employed.
Brillouin-zone sampling was restricted to the Baldereschi mean-value point \cite{Baldereschi1973_PRB}, which has been shown to provide accuracy comparable to that of a $4\times4\times4$ Monkhorst–Pack grid for liquid hydrogen at similar system sizes.
All AIMD simulations were performed in the canonical (NVT) ensemble using a time step of 0.5~fs and a Nosé–Hoover thermostat \cite{Nose1984_JCP,Hoover1985_PRA} for temperature control.
For each thermodynamic condition, the system was equilibrated for 1~ps, followed by a 1.5~ps production run.

Crystalline hydrogen structures were randomly generated using the \textsc{RandSpg} code for all cubic and hexagonal space groups, with 72 atoms per unit cell \cite{Avery2017_ComputPhysCommun}.
Geometry optimizations were performed at 100~GPa using the PBE functional within \textsc{VASP}.
Electrons were represented using PAW pseudopotentials with a plane-wave cutoff energy of 600~eV.
Reciprocal space was sampled using a $\Gamma$-centered $k$-point mesh with a maximum spacing of 0.15~\AA$^{-1}$.
Topological analysis of ELF isosurfaces was carried out using the \textsc{Critic2} package \cite{Otero2014_ComputPhysCommun}, and hydrogen-framework networking values were evaluated using the \textsc{TcEstime} framework \cite{Novoa2025_ChemSci}.

\section{Data Availability}
The code used in this study is publicly available at \url{https://github.com/July13210914/ELF-Prediction}. 
All scripts required to reproduce the training and evaluation procedures are included in the repository.
The data that support the findings of this study are available from the online repository https://www.lct.jussieu.fr/pagesperso/contrera/databaseELFprediction/ 

\section{Acknowledgments}
This work was supported by the Agence Nationale de la Recherche (ANR) under Grant No. ANR-22-CE50-0014 and by the ECOS-Sud program under Projects C17E09 and C21E06/ECOS210019. 

Computational resources were provided by GENCI under Projects No. A0160915069, A0160815101, and A0190915069, UKCP Archer at EPCC (EPSRC Grant No. EP/P022790/1) and by the Center for Computational Research at SUNY Buffalo (http://hdl.handle.net/10477/79221).

S.G. acknowledges support from MICIN PID2021-128005NB-C21 and RED2022-134890-T, Generalitat de Catalunya 2021SGR-633, and Universitat Rovira i Virgili 2025INTER-03 and 2023PFR-URV-00633. 
 M.M. acknowledges support from the ERC fellowship “Hecate”.  
E.Z. acknowledges the U.S Department of Energy, Office of Science, Fusion Energy Sciences funding the award entitled High Energy Density Quantum Matter, under Award No. DE-SC0020340. F.S. acknowledges support from the project PID2022-138327OB-I00, financed by the Ministerio de Ciencia e Innovación (MCIN)/Agencia Estatal de Investigación (AEI)/10.13039/501100011033/FEDER, UE.

\section{Conflict of interests}
The authors declare no conflict of interest.

\bibliography{reference}

\section{Supporting Information}
Supporting Information is available and includes: evaluation of the convolutional neural network (CNN) model; hyperparameter convergence tests; extended error analysis; and a supplementary table containing structure-wise statistics for randomly generated structures.

\begin{tocentry}
\includegraphics[width=\linewidth]{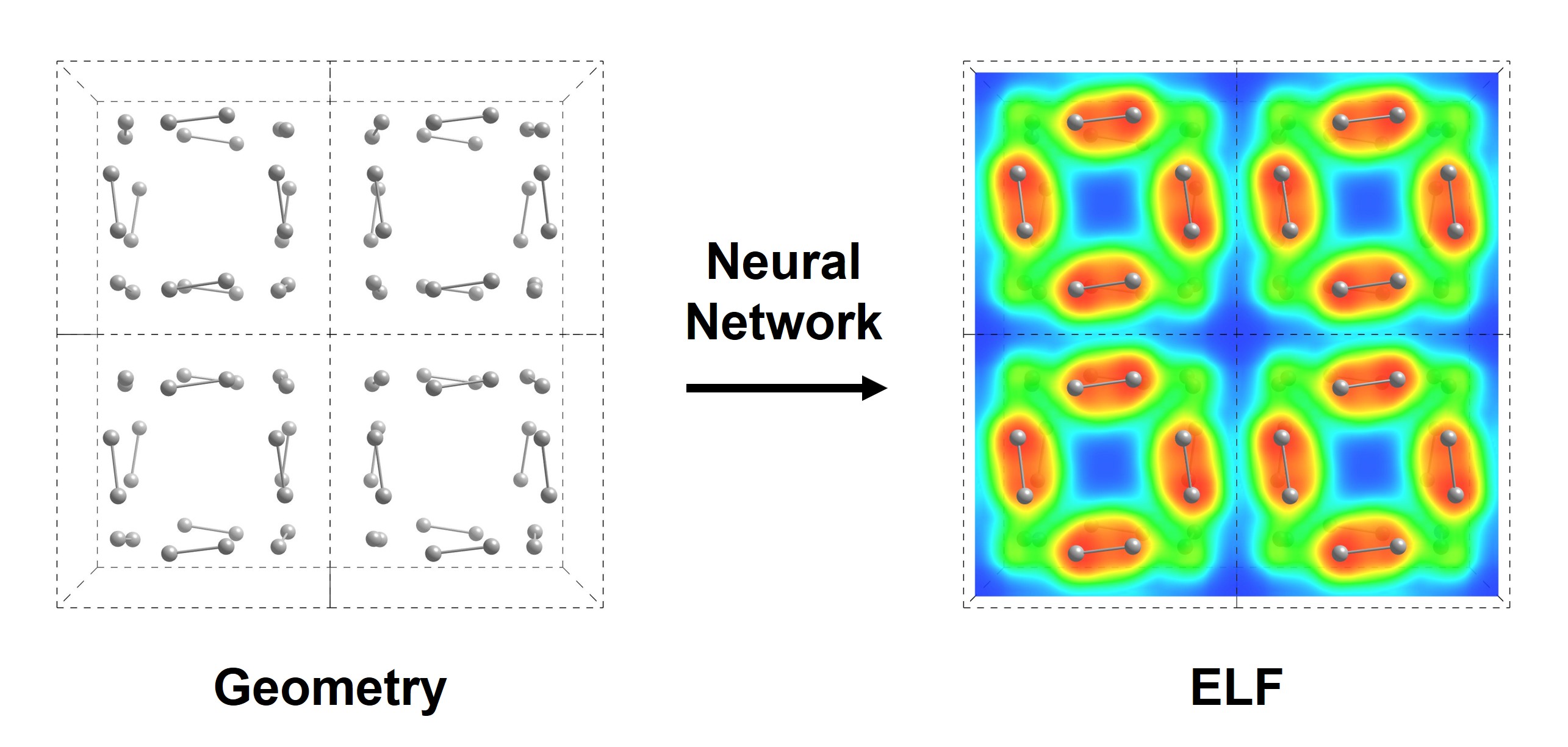}
We present a machine-learning framework that predicts the electron localization function (ELF) of dense hydrogen directly from atomic geometry, bypassing explicit electronic-structure calculations. Trained on ab initio data for fluid hydrogen across multiple pressures, the model achieves high accuracy and reveals pressure-dependent nonlocal contributions, while transferring robustly to crystalline hydrogen and preserving key ELF topological features.
\end{tocentry}

\end{document}